\documentclass[draft,jgrga]{AGUTeX}

  \usepackage[dvipdf]{graphicx}

  \setkeys{Gin}{draft=false}
\authorrunninghead{NIKOLAEVA ET AL.}

\titlerunninghead{DEPENDENCE OF CPCP SATURATION}

\begin{document}

%
%

\title{Dependence of the cross polar cap potential saturation on the type of solar wind streams 
 }
%
%

%
%




\authors{N. S. Nikolaeva, \altaffilmark{1}
Yu. I. Yermolaev, \altaffilmark{1} 
 I. G. Lodkina, \altaffilmark{1} 
}

\altaffiltext{1}{Space Plasma Physics Department, Space Research Institute, 
Russian Academy of Sciences, Profsoyuznaya 84/32, Moscow 117997, Russia. 
(nnikolae@iki.rssi.ru)}






%
%


\begin{abstract}

We compare of the cross polar cap potential (CPCP) saturation during magnetic storms induced by various types of the solar
wind drivers. By using the model of Siscoe-Hill \citep{Hilletal1976,Siscoeetal2002a,Siscoeetal2002b,Siscoeetal2004,Siscoe2011} we 
evaluate criteria of the  CPCP saturation during the main phases of 257 magnetic storms ($Dst_{min} \le -50$ nT) induced 
by the following types of the solar wind streams: 
magnetic clouds (MC), Ejecta, the compress region Sheath before MC ($Sh_{MC}$ ) and before Ejecta ($Sh_{E}$), 
corotating interaction regions (CIR) and indeterminate type (IND). 
Our analysis shows that occurrence rate of the CPCP saturation is higher for storms induced by ICME (13.2\%) than for storms driven by 
CIR (3.5\%) or by IND (3.5\%).The CPCP saturation was obtained more often for storms initiated by MC (25\%) than by Ejecta (2.9\%);
 it was obtained for 8.6\% of magnetic storms induced by sum of MC and Ejecta, and for 21.5\% magnetic storms induced by Sheath before 
them (sum of $Sh_{MC}$ and $Sh_{E}$). These results allow us to conclude that occurrence rate of the CPCP saturation at the main phase 
of magnetic storms depends on the type of the solar wind stream. 

\end{abstract}

%
%

%

\begin{article}

%
%

\section{Introduction}

As well known the main cause of geomagnetic storms is solar wind electric field $Ey = Vx \times Bz$, where Vx is 
radial component of solar wind velocity and Bz is the southward component of interplanetary magnetic field (IMF).
Solar wind includes various types of streams characterized by different behavior of strength 
and structure of IMF, density and velocity of solar wind, and these types of streams result in different forms of 
geomagnetic activity \citep{Boudouridisetal2004,BorovskyDenton2006,Huttunenetal2006,Pulkkinenetal2007,Yermolaevetal2007,
PlotnikovBarkova2007,Longdenetal2008,Turneretal2009,Despiraketal2011,Nikolaevaetal2011,Guoetal2011,Yermolaevetal2012}.

There are 5 geoeffective types/subtypes of the solar wind (SW): (1) Corotation Interaction Region (CIR), when 
high velocity stream of SW from coronal hole interacts with slow SW above the streamer belt; (2) Magnetic Clouds (MC), 
or well organized structures with enhanced IMF magnitude, large and smooth rotation of IMF vector over period $\sim 1$ day; 
low proton temperatures \citep{Burlagaetal1981}; (3) Ejecta, with less organized structure than MC; (4) Sheath or 
compression region before the leading edge of MC ($Sh_{MC}$); and (5) Sheath before Ejecta ($Sh_{E}$) 
(for example, see \citep{Yermolaevetal2009}).	

The cross polar cap potential saturation is one of differences between CME- and CIR- induced geomagnetic storms \citep{BorovskyDenton2006}.
It is known that potential across polar cap is increasing with growth of Ey. But sometimes its value does not change 
with increasing of Ey (i.e. it reaches the saturation threshold) under favorable solar wind conditions often associated with strong magnetic 
storms \citep{ReiffLuhmann1986,Russelletal2000,Russelletal2001,Nagatsuma2002,Shepherdetal2002,Oberetal2003,Hairstonetal2003,
Boudouridisetal2004,Hairstonetal2005,BorovskyDenton2006,Shepherd2007}.

The cross polar cap potential saturation is confirmed experimentally (for example, \citep{Nagatsuma2002,Shepherdetal2002,Hairstonetal2003, 
Oberetal2003}). Also this phenomena is agreed with MHD simulations \citep{Raederetal2001,Siscoeetal2002a,Merkineetal2003}. 
For an explanation of CPCP saturation it was proposed several models, although the physical mechanism is still debated 
\citep{Siscoeetal2002a,Siscoeetal2002b,Siscoeetal2004,KivelsonRidley2008,LavraudBorovsky2008,Borovskyetal2009,Gaoetal2013}.   

The authors \citep{Borovskyetal2009} compare several models for explanation of CPCP saturation dividing them 
into two types: ''reconnection models'' and ''postreconnection models''. The reconnection models explain the reduction of CPCP by 
reduction in the reconnection rate at the dayside of the magnetosphere, i.e., by reduction of SW-magnetosphere coupling 
\citep{Hilletal1976,Pudovkinetal1985,Raederetal2001,Siscoeetal2002a,Merkinetal2005a,Merkinetal2005b,RaederLu2005, 
Ridley2005,Hernandezetal2007}. The postreconnection models explain decreasing of CPCP by processes occurring after
the solar wind plasma reconnects with magnetosphere \citep{Wingleeetal2002,Siscoeetal2002b,Ridley2007,KivelsonRidley2008}. 
From these models the authors \citep{Borovskyetal2009} choose the  MHD-generator model \citep{KivelsonRidley2008} as 
the best one because it agree with results of global MHD modeling. 

The investigations (for example, \citep{LavraudBorovsky2008,Siscoe2011} show that CPCP predicted by \citep{Siscoeetal2002b} is 
similar one predicted by \citep{KivelsonRidley2008}. The similarities and differences between these two models were investigated 
in the work \citep{Gaoetal2013}. The authors compare mathematical formulas and predictions of both models with data measurements.
The results of the analysis show that both models predict similar saturation limits mathematically and give similar model 
predictions for CPCP value measured during time interval 1999--2009 \citep{Gaoetal2013}.  

Authors of the works \citep{Siscoeetal2002a,Siscoeetal2002b,Siscoeetal2004,Siscoe2011} on the basis of the hypothesis 
\citep{Hilletal1976}, have developed a theoretical model of coupling between the solar wind and the magnetosphere and ionosphere, 
which predicts the CPCP saturation. It occurs, when the region I current system 
generates a magnetic field which is approximately equal to dipole field at the dayside magnetopause 
\citep{Hilletal1976,Siscoeetal2002a,Siscoeetal2002b,Siscoeetal2004,Siscoe2011}.
Authors formulated the criterion of CPCP saturation which connects transpolar potential with the value of 
interplanetary electric field, solar wind dynamic pressure and ionospheric conductance 
\citep{Siscoeetal2002a,Siscoeetal2002b,Siscoeetal2004,Siscoe2011}.

According to numerous works 
\citep{Hilletal1976,Balanetal1993,Siscoeetal2002a,Siscoeetal2002b,Oberetal2003,Siscoeetal2004,Floydetal2005,BorovskyDenton2006}
the saturation of the polar cap potential occurs when a saturation parameter:
$$
		Q = Va \Sigma_{p} /806 = Va F10.7^{1/2}/1050 > 2\eqno   (1) 
$$
                                           
where $ Va $ is the Alfven velocity in the solar wind, $ \Sigma_{p} $ is the height-integrated Pederson conductivity of the ionosphere; 
according the work \citep{RobinsonVondrak1984} its value can be determined as  
$ \Sigma_{p} = 0.77 F10.7^{1/2} $ , where $F10.7$ is solar radio flux as proxy $ \Sigma_{p} $
(see details in papers by \citep{BorovskyDenton2006,LavraudBorovsky2008} and references therein). 

Using OMNI2 data set authors \citep{BorovskyDenton2006} obtained that the saturation of polar cap potential (i.e. $Q > 2$) was 
usually observed for CME-driven storms, but rarely observed for CIR-driven magnetic storms. It should be noted that in accordance 
with author's definition the CME-driven magnetic storms include all storms initiated by various interplanetary manifestations of CME: sheath, ejecta, 
and magnetic cloud \citep{BorovskyDenton2006}. Note that authors \citep{BorovskyDenton2006} used international sunspot number $Sn^{1/2}$ 
(with time resolution 1 month) as proxy $ \Sigma_{p} $. 

In contrast to previous papers we separately study magnetic storms induced by various components of CME manifestations. Also as proxy $ \Sigma_{p} $ we 
used the solar radio flux $F10.7$ which correlates with $Sn$ value, but gives more real values for $ \Sigma_{p} $ \citep{Oberetal2003}.  

The main aim of our work is an estimation of the CPCP saturation during the main phase of magnetic storms induced by 
different types of the solar wind streams which include CIR, and separately types of ICME such as magnetic clouds (MC), Ejecta, 
and Sheath before them ($Sh_{MC}$ and $Sh_{E}$, respectively). Separation of Sheath-storms on 2 types $Sh_{MC}$ and $Sh_{E}$ is partly 
justified by one of the results of the work \citep{Nikolaevaetal2011}. Magnetic storms induced by $Sh_{MC}$ have lower value 
of $Dst_{min}$ and higher value of $AE$ index.

In given work we analyze different types of magnetic storms (including their subtypes) in order to estimate what types/subtypes 
of SW more often lead to non-linear type of interaction with magnetosphere-ionosphere system (which manifests itself in CPCP saturation). 
In addition we used solar radio flux at 10.7 cm (with time resolution 1 day) as proxy $ \Sigma_{p} $. 

\section{Data} 

For our analysis we use OMNI data of interplanetary parameters and the ''The Catalog of Large-scale Solar Wind Phenomena 
during 1976--2000'' (site ftp://ftp.iki.rssi.ru/pub/omni/) \citep{KingPapitashvili2005,Yermolaevetal2009}. The method of identification 
of different types of SW on the basis of plasma and magnetic field data is described in detail in the work \citep{Yermolaevetal2009}. 
The technique of determination of connection between magnetic storms and their interplanetary drivers is the following. 
If the minimum of $Dst$ index lies in an interval of a type of solar wind streams or is observed within 1--2 hours after it we believe 
that the given storm has been generated by the given type of streams \citep{Yermolaevetal2010}.

To calculate the saturation parameter Q for different drivers we select 257 magnetic storms with $Dst \le -50$ nT and with full set of solar 
wind parameters needed for calculation of parameter according relation (1).
The solar wind data for calculation of Alfven velocity $Va$ 
and solar radio flux  $F10.7^{1/2}$, which used as proxy $ \Sigma_{p} $, were received from OMNI data base \citep{KingPapitashvili2005}.

The following types of the solar wind streams are sources of the magnetic storms: corotating interaction regions, CIR -- 56 magnetic storms; 
magnetic clouds, MC -- 24 magnetic storms; the compression regions ahead MC, $Sh_{MC}$ -- 5 events; Ejecta -- 69 events; the compression 
regions ahead Ejecta, $Sh_{E}$ -- 46 events, and indeterminate type IND (the sources which are impossible to determine because 
of data gap) -- 57 events. To compare results of this paper with previous results \citep{BorovskyDenton2006} we calculate similar 
parameters for sum of subtypes of ICME, magnetic clouds MC and Ejecta (MC+Ejecta) -- 93 magnetic storms, and compression 
regions before them Sheath ($Sh_{MC} + Sh_{E}$) -- 51 magnetic storms.

\section{Results}

Figure 1 shows the distribution of 257 magnetic storms with $Dst \le -50$ nT in dependence on type of the solar wind driver. 
We see that only 22\% of storms driven by CIR, but 56\% of all storms are driven by sum $MC+Ejecta+Sheath$ (including 36\% 
storms driven by sum $(MC+Ejecta)$ and ~20\% storms driven by Sheath ($Sh_{MC} + Sh_{E}$) ahead $(MC+Ejecta)$).

The results of evaluation of saturation parameter $Q$ and corresponding solar wind parameters: Alfven velocity $Va$ and the solar 
radio flux $F10.7^{1/2}$, which are included in the formula (1) for $Q$, are presented in Figure 2. In the Figure 2 the occurrence distribution 
of the polar cap saturation parameter Q is binned according type of solar wind drivers (top panel). In the middle and bottom 
panels in Figure 2 the Alfven velocity $Va$ and solar radio flux $F10.7^{1/2}$ are binned, respectively. 

The following designations are used for different types of magnetic storms (in Figure 2 a, c, e): thick blue line for CIR, thin blue line for IND, 
solid brown line for MC, dotted brown line for $Sh_{MC}$, solid red line for Ejecta, dotted red line for $Sh_{E}$, solid purple line for 
sum of $ MC+Ejecta$, and dotted purple line for sum $Sh_{MC} + Sh_{E}$, or Sheath. 
The right panels (b, d, f) in Figure 2 present the same data as in the left panels, but all magnetic storms are selected only into 3 main types 
of drivers as it was made in work by \citep{BorovskyDenton2006}: (1) CIR (thick grey line), (2) MC (thin black line), (i.e., CIR and MC repeat
 that on the left in Figure 1), and (3) ICME, which includes all of the interplanetary manifestations of CME: magnetic clouds (MC) 
and Ejecta, also the compression region Sheath (i.e. sum of $MC +  Ejecta + Sheath$). This type is close 
to CME-driven storms in paper \citep{BorovskyDenton2006} and below we compare them. The right panels (b, d, f) in Figure 2 
permit to compare our results with other works.

In Table 1 there are presented average and median values of Alfven velocity $Va$, parameter of saturation $Q$, and $F10.7^{1/2}$ 
(as proxy $ \Sigma_{p} $) for different types of SW drivers. We can see that MC- and $Sh_{MC}$ - storms have the highest values of $Q$, $Va$, 
$F10.7^{1/2}$; while CIR-, and IND-storms have the lowest ones with factors 
 ~2 , ~1.7 , ~1.3, respectively.

Median values of $Q$ depend on type of magnetic storms and change (in ~2.8-1.8 times) from maximal values 2.4 and 1.53 
(for $Sh_{MC}$- and MC- storms, respectively), to minimal values of 0.91 and 0.85 (for IND- and CIR- storms, respectively). The median 
value Q is higher for $Sh_{MC}$- storms than for MC-storms (factor ~1.6). In our sample of storms the factor between median values 
Q for MC- and CIR- storms is equal ~1.8  (against ~2.9 , in \citep{BorovskyDenton2006}. The median value of Q for CME-driven 
storms given by \citep{BorovskyDenton2006} is lower with factor ~1.4 in comparison with Q for ICME- driven storms (see Table 1
and Figure 2b). Such discrepancy may be explained by different events statistics in samples and by using $Sn^{1/2}$ as proxy $ \Sigma_{p} $.  

The median values Va are changing between maximum values 145 and 112 km/s for ShE- and MC-storms, respectively, and minimal 
values 70 and 78 km/s for IND-storms and CIR- and Ejecta- storms, respectively. For storms induced by (MC+Ejecta) and by 
compression region Sheath ($=Sh_{MC}+Sh_{E}$) the median values Va are very close (98.6 and 96 km/s, respectively). For storms induced 
by MC and $Sh_{MC}$ the factor between median values Va is ~1.3. The work 
\citep{BorovskyDenton2006}  contains the following median values of Va: 78 km/s for CIR-storms and 131 km/s for 
MC- storms and 95 km/s for CME- driven storms. These values are close to our results.   

The highest median values of solar radio flux $F10.7^{1/2}$ =16 are associated with $Sh_{MC}$- storms, the lowest values  $F10.7^{1/2}$ =12.5 and 
12.8 have storms induced by CIR and IND (factor 1.3). The magnetic storms induced by MC+Ejecta and by Sheath have 
equal median values $F10.7^{1/2}$ =13.5. But $Sh_{MC}$- storms have the median values $F10.7^{1/2}$ =16 larger than for MC-storms 
$F10.7^{1/2}$ =13.45 (with factor 1.2). In work \citep{BorovskyDenton2006}  there are presented following median values 
of $Sn^{1/2}$ (used as proxy $ \Sigma_{p} $): for CIR-storms $Sn^{1/2}$=4.8 (relative to our median value $F10.7^{1/2}$ =12.5), 
for MC- storms $Sn^{1/2}$=8.7 (our median value $F10.7^{1/2}$ =13.4), for CME-storms $Sn^{1/2}$=9.9 (our median 
value $F10.7^{1/2}$ =13.55). So the range of conductivity changing is equal 1.3 in our work (when solar radio flux $F10.7^{1/2}$ 
as proxy $ \Sigma_{p} $), in respect to factor 2 in work \citep{BorovskyDenton2006}, in which sunspot number $Sn^{1/2}$ was used 
as proxy conductivity. 

In Table 2 there are presented the number of magnetic storms driven by various types of SW for 3 levels of saturation parameter Q.
It is seen that high value of saturation parameter $Q>2$ was observed in 3.8 times more often for storms driven by ICME than by CIR 
and IND; also parameter $ Q>2$ is occurred in 8.6 times more often for MC- storms than for Ejecta- storms, and in 2.5 times more 
often for Sheath-storms than for (MC+Ejecta)- driven storms. 

Some decreasing of saturation parameter $Q>1.8$ (10\% decreasing of saturation parameter) leads to an increase number of storms 
driven by ICME (factor 1.2), mainly due to (MC+Ejecta)-driven storms than Sheath-driven storms (factor 1.37); also it leads 
to increasing number of Ejecta- and CIR-storms (with factors 2 and 2.6, respectively). The criterion $Q>1$ is performed 
for 2/3 of all ICME- storms versus 1/3 of CIR- and IND- storms, and for almost all $Sh_{MC}$- and MC- storms (~80\%).

\section{Discussion}

Obtained results not only confirm the conclusions of the work \citep{BorovskyDenton2006} that the storms driven by 
ICME(MC+ Ejecta+ Sheath) the most often satisfy criterion of CPCP saturation, but also we obtained indications that 
the most often the CPCP saturation is associated with magnetic storms driven by Sheath ($Sh_{MC}+Sh_{E}$) than by 
(ÌÑ+Ejecta) (21.5\% in comparison with 8.6\%, respectively). The occurrence rate of CPCP saturation 
for Sheath-driven storms is comparable with occurrence rate of saturation for MC-driven storms (21.5\% in comparison 
with 25\%, respectively). Thus during the main phase of magnetic storms the values of saturation parameter Q, Alfven velocity Va, and proxy 
Pederson conductivity $\Sigma_{p} \sim F10.7^{1/2}$ change in dependence on type of SW stream with the largest difference 
between them for CIR-driven storms and subtypes $Sh_{MC}$- and MC- driven storms. 

In contrast to paper by \cite{BorovskyDenton2006} we found saturation separately for different parts of ICME: 
MC and Ejecta, Sheath before MC and Ejecta.

The obtained results are not a surprise and may be explained by changing of SW parameters inside different types of SW streams 
which induced the magnetic storms. Also the occurrence rate of magnetic storms, induced by ICME, is higher near the maximum 
phase of solar activity when solar radio emission is stronger and ionospheric conductivity is higher. While the occurrence rate for 
CIR- driven magnetic storms is higher near the minimum phase of solar activity when solar radio emission is lower.

The Q values are dependent not only on variation of Va but also on $\Sigma_{p}$ variation. But on average contribution of Va is greater 
than contribution of $\Sigma_{p}$ (see Table 2). On average the contribution of Va in value of saturation parameter Q exceeds the contribution 
of the solar radio emission almost an order of magnitude (factor ~7-9). 

Figure 3 shows the saturation parameter Q versus Va and Q versus $F10.7^{1/2}$ for 4 types of SW streams CIR, MC, MC+Ejecta, 
Sheath($=Sh_{MC}+Sh_{E}$).

For all 4 types of SW the dependence of saturation parameter Q on Alfvenic velocity Va is linear with high coefficients of correlation 
(changes between r=0.92 for CIR-storms and r=0.97 for MC-storms). Coefficient of determination equals $r^2=0.94$ for MC-driven 
storms and $r^2$=0.84 for CIR-storms, that is about 94\% and 84\% variations of Q and Va are common for magnetic storms 
driven by MC and CIR, respectively. While linear dependence of the saturation parameter Q 
on the solar radio flux $F10.7^{1/2}$ 
(proxy $\Sigma_{p}$) is weaker (coefficients of correlation change between r=0.61 for CIR-storms and r=0.15 for MC-storms). 
Thus only 2\% and 5\% of the variations in Q and in $F10.7^{1/2}$ are common for MC- and MC+Ejecta- storms, respectively; 
and 20\% and 37\% of variations of both parameters are common for Sheath - and CIR- storms, respectively. The strong linear 
dependence of Q on Va with high values of correlation coefficients during magnetic storms driven by all types of SW may 
be explained by more large contribution of Va in value of parameter Q in comparison with ionospheric conductivity. 

As it is seen in Figure 3 a necessary condition for fulfilment of saturation criteria $Q>2$ is not only high Alfvenic velocity of SW
 ($Va>125-150$ km/s), that is high dayside reconnection rate, but also large ionospheric conductivity $\Sigma_{p}$ (range 
changing of solar radio flux $F10.7^{1/2} \sim 10-17$ corresponds to variation of conductivity $\Sigma_{p} \sim 7.7-13.1$ S). 
Contribution of each of these terms in the Q value depends on the type of SW stream. 
 
We can assume that 80\% of saturation can be explained by the processes external magnetosphere-ionosphere system \citep{Ridley2005}. 
The high Alfven velocity $Va$ means more efficient reconnection between interplanetary magnetic field at the dayside of magnetosphere. 
On the other hand $Va$ connected with Mach number $Ma ( \sim V/Va) $. The dependence of $Q$ versus $Ma$ (not presented here) show that 
criteria of saturation ($Q>2$) corresponds to the low values of Mach number ($Ìà<4.5$) for all types of magnetic storms.  

It should be noted that we used in our calculations the solar radio flux  $F10.7^{1/2}$ as proxy integrated Pederson conductivity $\Sigma_{p}$. 
The real system of field aligned currents also includes currents zone 2, but usually it not presented in MHD models 
(e.g., \citep{Raederetal1998}). Further investigations are required.

\section{Conclusions}

By using the model of Siscoe-Hill \citep{Hilletal1976,Siscoeetal2002a,Siscoeetal2002b,Siscoeetal2004,Siscoe2011} we 
evaluate criteria of the  CPCP saturation (parameter saturation $ Q =  Va F10.7^{1/2}/1050 > 2 $) during the main phases of 257 magnetic storms
 ($Dst_{min} \le -50$ nT) induced by the following types of the solar wind streams: 
corotating interaction regions, CIR -- 56 magnetic storms; 
magnetic clouds, MC -- 24 magnetic storms; the compression regions ahead MC, $Sh_{MC}$ -- 5 events; Ejecta -- 69 events; the compression 
regions ahead Ejecta, $Sh_{E}$ -- 46 events, and indeterminate type IND (the sources which are impossible to determine because 
of data gap) -- 57 events. Also we calculate similar parameters for sum of subtypes of ICME, magnetic clouds MC 
and Ejecta (MC+Ejecta) -- 93 magnetic storms, and compression 
regions before them Sheath ($Sh_{MC} + Sh_{E}$) -- 51 magnetic storms.

We 
obtained and analyzed the occurrence distribution of saturation parameter $Q$ values, of the Alfven velocity $Va$ and of solar radio 
flux $F10.7^{1/2}$ (as proxy height-integrated Pederson conductivity $\Sigma_{p}$)
according type of solar wind drivers. 

The median values of $Q$ depend on type of magnetic storms and change in $\sim$ 2.8-1.8 times between maximal values 
for $Sh_{MC}$- and MC- storms and minimal values for CIR- storms. 
The median value Q is higher in  $\sim$ 1.6 times for $Sh_{MC}$- storms than for MC-storms.

The median values of $Va$ are changing in 1.4-1.8 times between maximum values for ShE- and MC-storms
and minimal values for CIR- and Ejecta- storms, respectively.
For storms induced by MC and $Sh_{MC}$ the factor between median values Va is $\sim$ 1.3.

The median values of solar radio flux $F10.7^{1/2}$ change in 1.3 times between maximum values for $Sh_{MC}$- storms
and minimal values  $F10.7^{1/2}$  for CIR-storms.
The median values $F10.7^{1/2}$  are larger in $\sim$ 1.2 times for $Sh_{MC}$- storms than for MC-storms.  

Thus we obtained that during the main phase of magnetic storms the values of saturation parameter $Q$, Alfven velocity $Va$, and proxy 
Pederson conductivity $\Sigma_{p} \sim F10.7^{1/2}$ change in dependence on type of SW stream with the largest difference 
between them for CIR-driven storms and subtypes $Sh_{MC}$- and MC- driven storms. 

The saturation parameters $Q$ values are dependent on  variations of  both parameters as Alfvenic velocity $Va$ as ionospheric conductivity
$\Sigma_{p}$. But on average the contribution of $Va$ in value of saturation parameter $Q$ exceeds in $\sim$ 7-9 times the contribution of  $\Sigma_{p}$ variation
of the solar radio emission $F10.7^{1/2} ( \sim \Sigma_{p}$). 

Our analysis allows us to make following main conclusions.

1) On the main phase of magnetic storms the CPCP saturation depends on type of the solar wind stream induced the magnetic storm.
 
2) The saturation criterion ($Q>2$) of the CPCP is performing mainly for strong magnetic storms initiated by ICME(MC+Ejecta+Sheath)  
    (~13.2\% storms), and in $\sim 3.5$ times rarely for CIR- and IND- storms (3.5\%).

3) Most often saturation criterion ($Q>2$) of cross polar cap potential is satisfied for storms driven by MC (25\%) than by Ejecta (2.9\%); 

4) The saturation ($Q>2$) of cross polar cap potential in  2.5 times more often is satisfied for Sheath- storms (21.5\%) than for storms 
   driven by sum of MC+Ejecta (8.6\%); 

5) Decreasing of saturation level on 10\% ($Q>1.8$) increases the number of ICME-storms with the CPCP saturation to 20\% 
    (by 40\% due to storms driven by sum of MC+Ejecta and by 9\% due to storms driven by Sheath).


%
%
%
%
%
%
%

\begin{acknowledgments}
The authors are grateful for the possibility of using the OMNI database. The OMNI data were obtained from 
the GSFC/SPDF OMNIWeb on the site http://omniweb.gsfc.nasa.gov. 
This work was supported by the Russian Foundation for Basic Research (RFBR), project 13-02-00158a, and 
by the Program 22 of Presidium of Russian Academy of Sciences.

\end{acknowledgments}

\end{article}


%
%

%
%
%
%
%


\begin{figure}
\noindent\includegraphics[width=12cm]{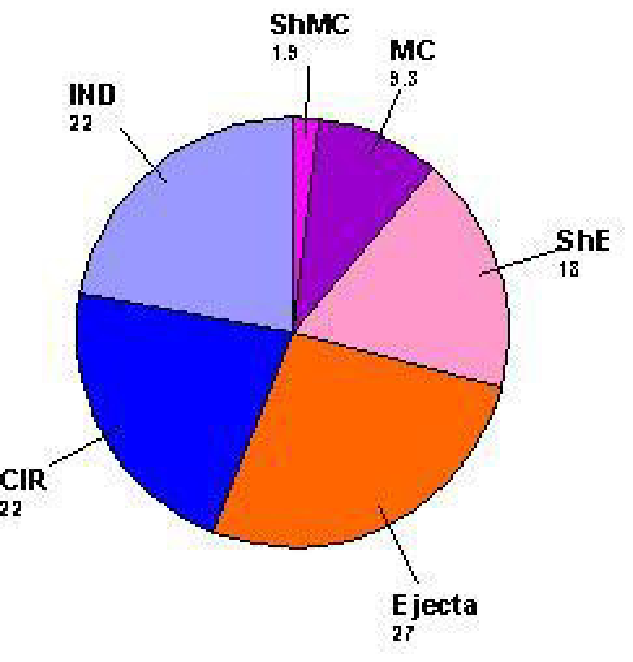}
\caption{The distribution of magnetic storms with $Dst \le -50$ nT in dependence on type of the solar wind driver (in \%). 
}
\end{figure}


\begin{figure}
\noindent\includegraphics[width=12cm]{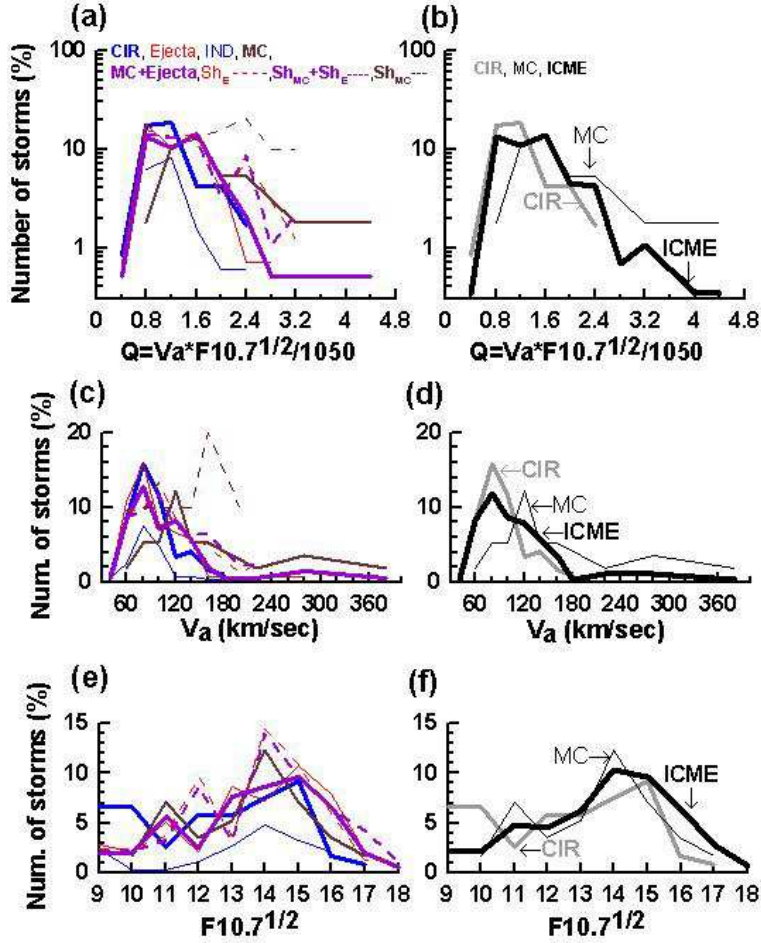}
\caption{Distributions of $Q$, $Va$, $F10.7^{1/2}$ for different types/subtypes of SW drivers. 
Panels (b, d, f) give the same distributions only 
for 3 types of SW drivers: CIR, MC, and ICME (= MC + Ejecta + Sheath).
}
\end{figure}


\begin{figure}
\noindent\includegraphics[width=12cm]{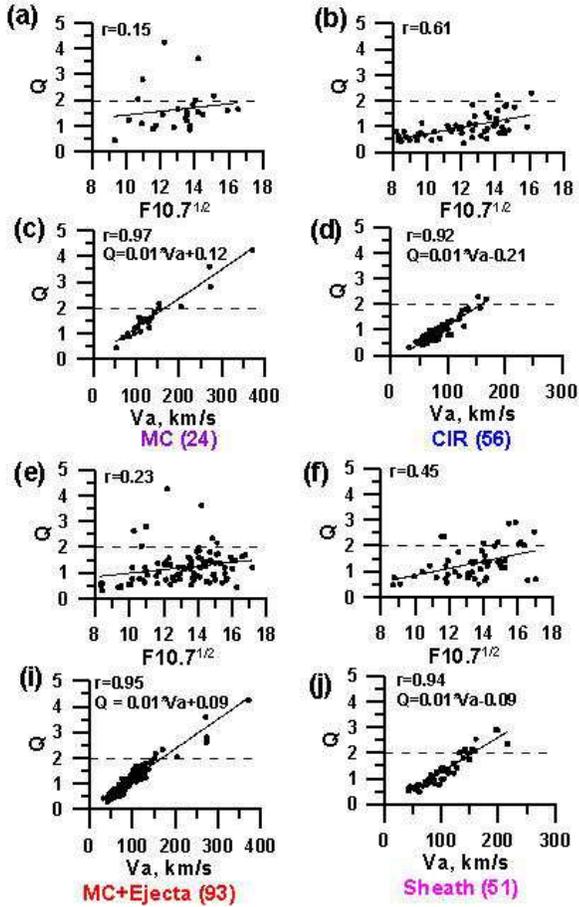}
\caption{A saturation parameter $Q$ versus $Va$ (c, d, i, j) and $Q$ versus $F10.7^{1/2}$ (a, b, e, f) for 4 types of magnetic storms induced 
by SW streams: MC (a, c), CIR (b, d), sum of (MC+Ejecta) (e, i), and Sheath ($Sh_{MC}+Sh_{E}$) before (MC+Ejecta) (f, j).
}
\end{figure}

%


\begin{table}
\caption{Average and median values of $Va$, $Q$, $F10.7^{1/2 }$ for magnetic storms induced by different types of SW.
}
\centering
\begin{tabular}{l ccccccc}
\hline
 Type of SW                             & Number of &  Va       &    Va    & $F10.7^{1/2 }$ & $F10.7^{1/2 }$ &  $Q$ &  $Q$ \\
                                                &    storms   &  Average & Median & Average & Median &  Average & Median \\
\hline 
  MC                                           & 24          &  134.9    &  112       & 12.98   & 13.45     &  1.65      & 1.53   \\
  CIR                                           & 56          &    84.2   &    78       & 12        &  12.5      &  0.98      & 0.85   \\
  Ejecta                                       & 69          &    85.8   &    78       & 13.18   &  13.56    &  1.08      & 0.96   \\
  $Sh_{E}$                                   & 46          &    99.7    &    95       & 13.1     &  13.5      &  1.2       & 1.2     \\
  $Sh_{MC}$                                &  5          &   140.7   &   145       & 15.6    &  16         &  2.12      & 2.4    \\
  Sheath ($Sh_{MC} + Sh_{E}$)     & 51         &   104      &    96        & 13.3    &  13.5       &  1.33      & 1.2    \\
  MC+Ejecta                                 & 93         &    98.5   &    98.5     & 13.1    &  13.5       &  1.2        & 1.2    \\
  IND                                            & 57         &    76.4   &    70        & 12.8    &  13.5       &  0.94      & 0.91   \\
  ICME (MC+Ejecta+Sheath)         & 144       &   100.3   &    93.5     & 13.2     &  13.55     &  1.26       & 1.2     \\
\hline
\end{tabular}
\end{table}


\begin{table}
\caption{The number of magnetic storms driven by various types of SW for 3 levels of saturation parameter $Q$.
}
\centering
\begin{tabular}{l cccc}
\hline
 Type of SW                            & Number of  & $N$ with $Q>2$ ,  & $N$ with $Q>1.8$,  & $N$ with $Q>1$ , \\
                                               &    storms     &   (\% of storms)     &  (\% of storms)       &  (\% of storms)  \\
\hline 
  MC                                                  & 24          & 6 (25\%)         & 7 (29\%)             & 20 (83\%)     \\
  CIR                                                  & 56          & 2 (3.5\%)        & 5 (9\%)              & 21 (37.5\%)    \\
  Ejecta                                              & 69          & 2 (2.9\%)        & 4 (5.8\%)           & 32 (46.4\%)    \\
  $Sh_{E}$                                          & 46          & 7 (15\%)         & 8 (17.4\%)         & 27 (58.7\%)   \\
  $Sh_{MC}$                                       &  5          & 4 (80\%)         & 4 (80\%)            & 4 (80\%)      \\
  Sheath ($Sh_{MC} + Sh_{E}$)            & 51         & 11 (21.5\%)     & 12 (23.5\%)       & 31 (61\%)    \\
  MC+Ejecta                                        & 93         & 8 (8.6\%)         & 11 (11.8\%)       & 52 (56\%)    \\
  IND                                                   & 57         & 2 (3.5\%)         & 2 (3.5\%)           & 19 (33\%)    \\
  ICME (MC+Ejecta+Sheath)                & 144       & 19 (13.2\%)      & 23 (16\%)         & 83 (57.6\%)  \\
\hline
\end{tabular}
\end{table}

\end{document}